# Controlling Multipolar Surface Plasmon Excitation through the Azimuthal Phase Structure of Electron Vortex Beams.

Daniel Ugarte[†,‡,*], Caterina Ducati[†].

[†]Department of Materials Science and Metallurgy, University of Cambridge, Cambridge CB3 0FS, United Kingdom.
[‡]Intituto de Física "Gleb Wataghin", Universidade Estadual de Campinas-UNICAMP, 13083-859 Campinas SP, Brazil.

**ABSTRACT:** We have theoretically studied how the azimuthal phase structure of an electron vortex beam excites surface plasmons on metal particles of different geometries as observed in electron energy loss spectroscopy. To do so, we have developed a semi-classical approximation combining an azimuthal phase factor and the dielectric formalism. Our results indicate that the vortex beam order may be used to modify and control surface plasmon multipole excitation in nanoparticles. In favorable cases, specific plasmon modes can even attain enhancement factor of several orders of magnitude. Since, electron vortex beams interact with particles mostly through interference effects due to azimuthal symmetries, i.e. in the plane perpendicular to the electron beam, anisotropy information (longitudinal and transversal) of the sample can be derived in EELS studies by comparing non-vortex and vortex beam measurements.

The exponential growth of nanoscience has stimulated an astonishing control of the chemical and physical properties of nanoparticles targeting technological issues in fields so different such as material science, catalysis, electronics, biology, medicine, etc. Transmission electron microscopy methods have in parallel increased their already remarkable spatial resolution and analytical capabilities; in particular, the increase in energy resolution has allowed electron energy loss spectroscopy (EELS) to begin to probe optical transitions. The combination of these two factors has allowed the nanometric spatial resolution study of electronic excitation of nanoparticles (NPs), a very active field of research, in particular addressing collective oscillation of valence electrons, called surface plasmons (SPs), which are strongly dependent on particle shape and electronic properties[1,2]. The mapping of SP excitation represents one of the essential applications of advanced Scanning Transmission Electron Microscopy (STEM) instruments. In spite of recent progress in physical understanding, the comparison between electron and light-based spectroscopies is not always straightforward, because electrons do not obey such tight energy and momentum selection rules[1-3], then they can excite several modes, , which are not accessible by light scattering experiments, the so called dark SP modes.

The recent generation of electron vortex beam carrying well-defined orbital angular momentum (OAM)[4-6] has stimulated several studies aiming to exploit these beams to study core level magnetic transitions[7-9] and magnetic plasmons[10]. Also, the interaction of vortex beams with helical molecules, nanoparticles and crystals has been studied theoretically in order to analyse the possibility of detecting chirality in nanostructures by momentum-resolved EELS spectra or electron diffraction[11,12]. Another important aspect of electron vortex beams, not fully exploited yet, is the azimuthal phase variation[13] which can be used to manipulate the interaction with the sample. We must note that Hickmann *et al.*[14] have experimentally demonstrated that a light beam carrying angular momentum generates a truncated triangular optical lattice in the far field when diffracted by a triangular aperture. These observations were qualitatively explained by describing the triangular aperture as three narrow slits considering a phase shift generated by the azimuthal phase structure of the vortex beam. Recently, a theoretical study has analysed how the phase spatial structure of vortex beams can be used to focus the beam to subwavelength size using a metallic nanoscale resonant optical antenna[15]. These reports are very instructive because they associate the spatial distribution of the azimuthal phase in a vortex beam with the in-plane rotational symmetry of the target.

It is intellectually tempting to extend this concept to atomic orbital wave functions and stimulate or reduce selected electronic transitions in spectroscopy. However vortex beams have a vanishing intensity at the beam centre (as emphasize in Ref. [15], a "donut-cross-section"), which makes it extremely difficult to get a stronger effect by focusing the beam. For vortex electron beams, the central zero intensity is not so restrictive, because of the long range Coulomb interaction between electron beam and sample (ex. aloof spectroscopy [16-18]). The simplest picture for such experiments would be a metal nanoparti-cle located in the central region of the vortex beam where the intensity vanishes. It is well known that surface plasmon modes in nanoparticles are mainly determined by size and morphology, so this represents an excellent case to study the coupling between the NP electronic excitations (surface plasmons) and a vortex beam azimuthal phase structure.

Here, we have theoretically studied how the azimuthal phase structure of an electron vortex beam excites surface plasmons on metal nanoparticles of different geometries as observed in electron energy loss spectroscopy (EELS). To do so, we have developed a semi-classical approximation combining an azimuthal phase factor and the dielectric formalism based on the Boundary Element Method (BEM[19,20]). Our results indicate that the vortex beam order may be used to select or manipulate multipole excitation in nanoparticles, and in favorable cases even attain an enhancement factor of several orders of magnitude. Our study has revealed that vortex beams interact with particles mostly through symmetry effects in the plane perpendicular to the electron beam. In this way, anisotropy information (longitudinal and transversal) of the sample can be derived in EELS studies by comparing non-vortex and vortex beam measurements.

Among the different approaches to calculate SP excitation by an electron beam, the boundary element method (BEM), originally proposed by Garcia de Abajo *et al.*[19,20], is one of the most popular procedures. The BEM method can be easily implemented for homogeneous particles (described by a local dielectric response $\varepsilon_1(\omega)$, $\omega$ representing the energy) of arbitrary shape embedded in a uniform dielectric medium ($\varepsilon_2(\omega)$). The calculations require the knowledge of the surface charge distribution $\sigma(s)$ at an assumed abrupt interface between the two dielectric media. Using the quasi-static approximation, the surface charge $\sigma(s)$ is derived from Poisson´s equations for a fast electron travelling close to the dielectric particle. The BEM surface charge density is obtained from a self-consistent integral equation in the energy domain (see details in Ref. [19], a brief description is included in the Supporting Information).

It is important to emphasize that the surface charge distribution can be built by the weighted contribution of different interface surface modes $\sigma^i(s, \omega)$, which are independent of energy and only depend on the nanoparticle morphology[19-22]. These so-called geometrical eigenfunctions describe self-sustained surface charge oscillations modes (or surface plasmons), which are obtained from an eigenvalue ($\lambda_i$) integral equation. Ouyang and Isaacson [23] have shown that the $\lambda_i$ factors are real, and the modes form a complete basis set that satisfies a particular (bi-)orthogonality property. The final surface charge density is in fact the weighted sum over the surface charge modes, which allows the calculation of the boundary polarization potential.

The energy loss is calculated by considering the dissipative work performed on the fast electron by the electric field originated from the boundary potential. Without loss of generality, we may consider an electron travelling along a rectilinear path, $r_e = r_0 + v\,t$, where $v = (0,0,-|v|)$ and $r_0$ is perpendicular to the velocity $v$. Finally, we consider that the medium around the particle is vacuum ($\varepsilon_2(\omega) \equiv 1$). When the electron travels outside the particle volume (hereafter we will always assume this configuration), the energy loss probability for each surface mode $i$ results[21,22]:

$$P^i(\boldsymbol{r_0}, \omega) = \frac{1}{(1+\lambda_i)\pi^2 v} \quad Im\{-g_i(\omega)\} \quad \text{X}$$

$$\left\| \int d\boldsymbol{s}\, \sigma^i(\boldsymbol{s}) e^{-i\frac{s^{\parallel}\omega}{v}} \; K_0\left(\frac{\omega|r_0 - s^{\perp}|}{v}\right) \right\|^2$$

where $\quad g_i(\omega) = \frac{2}{\varepsilon_1(\omega)(1+\lambda_i) + \varepsilon_2(\omega)(1-\lambda_i)}$

The indexes $\perp$ ($\parallel$) represents the vector component perdendicular (parallel) to the electron trajectory, and $K_0$ is the modified Bessel function of order 0. This expression is very instructive because it clearly shows that the main energy dependence is contained in the $g_i(\omega)$ factor, and geometrical aspects in the squared integral factor.

A strengh of this formalism is that it can be easily implemented numerically by discretizing the surface modes[20]; in fact, the final enegy loss probability results from the interaction between point charges ($\sigma^i(s)\,ds$ and the electron, mathematically included in the $K_0$ function). In order to study the excitation of SP modes by electron vortex beams, we must find a physical description for an electron vortex beam or wave that maintains this point-charge-point-charge interaction profile. Mohammmadi *et al.*[10] have considered an electron travelling along a helical trajectory to generate an electron beam with angular momentum with the aim of studying magnetic responses. In contrast, our aim is to find a description where the azimuthal phase structure of the vortex beam could be taken into account within a quasi-static electromagnetic calculation. As mentioned previously, since vortex beams present a central region with vanishing intensity, we may approximate the beam cross section as being an infinitively thin ring of radius $R_0$ (see Figure 1). In analogy to the numerical implementation in the BEM approach, we discretize the vortex beam as an ensemble of non-interacting linear electron trajectories. Then, for each position along the azimuthal $\varphi$ angle ($\boldsymbol{r}_{0,k} = \boldsymbol{r}_0(\varphi_k), |\boldsymbol{r}_{0,k}| = R_0$ ), we add an additional spatial phase term $exp(in\varphi)$ in order to generate the vortex azimuthal phase structure of order $n$ ($n$ integer):

$$\rho(\boldsymbol{r}) = A \sum_k e^{i\,n\,\varphi_k}\, \delta(\boldsymbol{r} - \boldsymbol{r}_{0,k})$$

where $A$ is a normalization constant derived from the volume integral $\int \rho^*\rho\; dV = 1$

This expression represents the perpendicular dependence of a vortex wave[13], where the azimuthal phase factor is associated with an electron beam constrained spatially to an infinitely thin ring of delta functions. Subsequently, we calculate the energy loss adding the contribution of each electron beam considering the vortex phase factor. The calculation of the energy loss considering this extra summation does not add major ancillary difficulties to the BEM dielectric formalism (see details in the Supporting Information). Within this pseudo-static calculation, the dissipation work results from the electric field due to the boundary potential evaluated at the electron position. As the surface charge density is derived considering the whole ring-shaped beam, the boundary potential will certainly result in a weighed sum including different spatial phase factors. Without any further approximation and following the typical result of the BEM procedure for the donut beam, the energy loss associated with the surface charge mode $\sigma^i$ results (see details in the Supporting Information):

$$P_n^i(\boldsymbol{r_0}, \omega) = \frac{4}{\pi v^2} \frac{(2+\lambda_i)}{(1+\lambda_i)} \; Im\{-g_i(\omega)\} \; \times$$

$$\left\| \sum_k e^{-i n \varphi_k} \int d\boldsymbol{s} \; \sigma^i(\boldsymbol{s}) e^{-i\frac{s^{\|}\omega}{v}} \; K_0\left(\frac{\omega \, |\boldsymbol{r_0}-\boldsymbol{s}^{\perp}|}{v}\right) \right\|^2$$

These expression is almost identical to the one derived for an individual electron trajectory[21,22]. Each azimuthal position of the discretized electron beam contributes independently to the energy loss through the surface charge integral. Subsequently, these individual terms (each electron position in the beam cross section) are added taking into account the spatial phase factor, and then the square modulus is evaluated, as expected for a phase effect is a quantum description. In this way, the vortex phase structure will directly interact with the azimuthal symmetry of SP modes described by $\sigma_i(\boldsymbol{s})$.

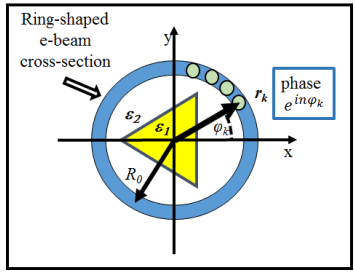

Figure 1. Schematic representation of the configuration used to calculate the interaction between plasmon modes and a vortex electron beam of order *n*. The donut-cross-section of the vortex beam is described as an infinitively thin ring of radius $R_0$ and, then we perform the calculation by adding the contribution of a series of parallel non-interacting electron beams along this ring but for each azimuthal angle $\varphi_k$, we assign the typical vortex phase term $e^{in\varphi_k}$ to that electron trajectory. The nanoparticle is described by a dielectric function $\varepsilon_1(\omega)$ immersed in a surrounding medium described by $\varepsilon_2(\omega)$.

In the following sections, we will apply this approach to calculate the energy losses associated with SP modes for particles of different morphologies located at the centre of the vortex beam (see Figure 1). We will study silver nanoparticles immersed in vacuum ($\varepsilon_2(\omega) = 1$), where Ag dielectric properties were taken from experimental values[24]. Considering the vortex index *n*, we will analyse rather low values (1, 2, 3,…) associated with dipolar (*D*), quadrupolar (*Q*), hexapolar (*H*), etc., rotational symmetries. We will also consider the vortex order *n=0* to represent electron beams without rotational symmetry. This case is interesting both because it represents the energy loss spectrum of a nanoparticle obtained using a standard extended (non-focused) electron beam, and it provides a reference value for SP excitation probabilities in relation to experimental measurements. We have used between 500-1000 elements for BEM calculations, and 50-100 electron positions along the donut beam circumference. For all energy loss calculations presented here, we have considered an electron energy of 100 keV.

As a first case study, we have addressed the SP excitation on spherical NPs. Although, this highly symmetrical morphology does not seem suited to analyse azimuthal geometrical effect, the surface charge modes are the well-known spherical harmonics $Y_l^m$ functions, which display well-defined polar and azimuthal dependence on the *l* and *m* indexes respectively. It is also important to emphasize that spherical harmonics allow many analytical calculations, as for example eigenvalues ($-1/(2\,l+1)$) and normalization factors ($\sqrt{(2\,l+1)/(4\,\pi\,a^3)}$ , *a* radius of the sphere), which can be very useful to test numerical software implementations.

When performing a low loss EELS study inside an electron microscope, multipolar SP excitations are dominated by the dipolar mode (*l=1*) for spherical nanoparticles, and higher multipoles have a much lower excitation probability. Among the dipolar modes, the energy loss is due to $Y_1^0$, because the induced electric dipole is parallel to electron trajectory (along z axis here). When considering a vortex beam with rotational phase structure ($n \geq 1$), this mode contribution vanishes (rotationally symmetric along z) and, the in-plane dipolar modes ($Y_1^{\pm 1}$) lying perpendicular to the electron trajectory become excited. This represent a very important property, because the use of vortex electron beams allows the study of excitations whose electric field lie mainly perpendicular to the electron trajectory and, show very weak contribution in standard TEM spectroscopic studies.

In analogy to hybridization of atomic orbitals (*sp*, *sp²*, *sp³*, etc.), spherical harmonics can be combined to get a desired surface charge mode showing well-defined azimuthal geometry. For example, we could generate an hybridized mode by the sum $h_l = (Y_l^l + (-1)^m Y_l^{-l})$; this yields a SP mode that is real and with an azimuthal modulation $cos(l\varphi)$ . These $h_l$ modes have a well-defined eigenvalue as $\lambda_i$ depends exclusively on the *l* index and, they represent an ideal case to analyse the effect of vortex beams on SP excitation.

The calculations show that for each vortex beam the only excited $h_l$ SP mode matches exactly the vortex order *n* and, all other modes display negligible contribution to the energy loss (this effect can be visualized in the progressive blue shift of the plasmon peak for increasing vortex order, Figure 1a). Concerning the intensity expected, the use of vortex beams enhance the mode excitation by several orders of magnitude for the first vortex beams (attaining almost a 10⁴ enhancement factor for the octupole $l=4$), and diminishes for higher $n$ values (Figure 1 b). We attribute this decrease to the increasing role of the polar angle in the spherical harmonics and, also the $s^{\parallel}$ dependent dephasing term inside the surface charge integrals. Our theoretical model predicts that pure quadrupolar or hexapolar modes can be excited and their intensities should be just 1 or 2 orders of magnitude lower than the measured values using non vortex beams (Figure 1b).

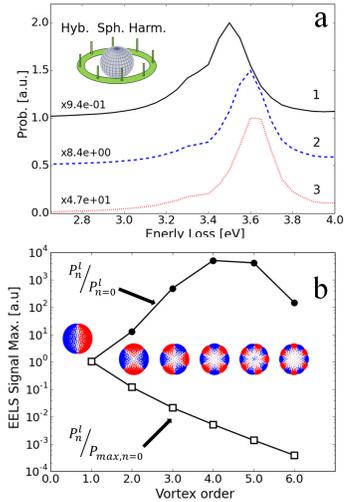

Figure 2. Calculated energy loss probability for hybridized sphere SP modes ($h_l$) with well defined azimuthal symmetry obtained by combining spherical harmonics of $m=\pm l$ order ($h_l = Y_l^i + (-1)^m Y_l^{-l}$). The calculation considers a silver sphere of radius 5 nm located at the centre of a vortex donut beam of radius 8 nm and modes up to $l = 6$ were included in the calculations. a) Predicted spectra for different vortex order ($n = 1$-$3$); the intensity has been normalized in relation to the maximum signal predicted for the donut-beam with order $n=0$ and they have been shifted vertically by 0.5 for an easily visualization. b) In order to estimate potential symmetry-based enhancement effects, we compare the excitation of each $h_l$ surface mode for each vortex order ($P_n^l$) in relation to predicted probability of the same mode for a vortex order $n=0$ ($P_{n=0}^l$; the values are indicated by black dots, the lines are just a guide for the eyes). This figure also contains the intensity evolution of intensity of the $h_l$ modes with vortex order normalized by the maximum of the spectra predicted for vortex $n=0$ in order to get an estimation of the experimental observability (indicated by white squares).

Briefly, the results based on a static dielectric calculation demonstrate a clear resonance effect between vortex order and SP mode rotational symmetry. Also, we must keep in mind that experiments deal with a dynamic situation much closer to a forced oscillator configuration (the driving force being the vortex beam). Then, the hybrids $h_{l=n}$ modes should spontaneously arise on the spherical nanoparticles and dominate the energy absorption. From an experimental point of view, plasmon peaks are rather broad and the energy distance is rather small in spherical NPs, then, it would extremely difficult to measure multipolar contributions for spherical nanoparticles. However, this may be performed considering other nanoparticle morphologies such as triangles or disks[25-28].

In recent years, the number of EELS-based surface plasmon studies has increased significantly by the combination of different factors, including chemical synthesis (shape control), spatial resolution (electron optic progress, narrow beams), TEM equipped with electron monochomatized beams, among others. Noble metal nanoparticles of different shapes (wires, triangles, disks, arbitrary shapes) have revealed several new phenomena. In particular, high-aspect ratio triangles or disk particles display SP modes which show very distinct SP modes with well separated SP peaks on the energy scale. In addition, these particles are almost ideal TEM samples, thin and easy to observe in imaging modes. As these systems are rather thin, surface modes are symmetrical (or surface charge is identical in both surfaces (perpendicular to the small thickness direction)[26]. Then the SP modes are defined by the geometrical distribution on the extended and usually planar surface. In a similar way, narrow wires develop modes that are mostly Fabry-Perot reflections along the wire length[29].

The study of disk-shaped metal particles prepared by lithographic methods have allowed the identification of surface modes defined by azimuthal symmetry (dipolar, quadrupolar, hexapolar, etc.)[28]. Figure 3 shows the energy loss probability for disk-shaped NPs located at the centre of vortex beams. In analogy to the spherical case, there is clear matching between vortex order and SP azimuthal index, but as the SP peaks are very distinct spectral features, the modes are easily identifiable in the calculated energy loss spectra. As the particles are thin and flat with almost negligible 3D structure, the dephasing factor ($s^{\parallel}$ dependent exponential inside the surface charge integrals) has little effect and several orders of magnitude enhancement is observed (Figure 3b), in such a way that the dipolar peak is about ~1000 times higher than the maximum of intensity expected for the vortex $n= 0$ spectra. The reinforcement is lower for higher modes ~100 and ~10 for $Q$ and $H$ respectively, indicating that this effect should be easily observed experimentally.

Disc-shaped nanoparticles show such enhancement in SP excitation by electron vortex beams due to the regular shape; we must remind that our calculations have considered a perfectly matched configuration, with the disk nanoparticle located at centre of a donut electron beam. Another noble metal NP morphology intensively used in EELS-based SP studies is the triangular shape[25,26,28]. The excitation of SPs on this kind of particles is fairly well understood, however the interaction donut beam-triangle is much more complex than for discs. Firstly the particle shape imposes several geometrical constraints, and although the relevant modes are usually described as dipolar, quadrupolar or hexapolar[28], their actual geometrical pattern is far from a well-defined rotational symmetry; please note that these denomination somewhat represent roughly the charge distribution (see Figure 4)[28]. By studying SP excitations for particles with shapes between a disk and a triangle, Scmidt et al.[28] have shown that we should expect two modes for dipole ($D_1$, $D_2$) and quadrupole ($Q_1$, $Q_2$) orders; however, one single mode arises with 3-fold ($H$) symmetry. Spatially resolved studies have shown that dipolar modes are strongly excited when the electron trajectory is close to the triangle apexes and the quadrupolar modes are mostly exited for electron trajectories close to triangle sides[25,26]. Hexapolar modes would be naturally expected due to the particle shape, however their cross-section seems to be rather low and difficult to detect in experiments. When using a focussed narrow beam experiments we expect a strong localization of the polarization charges close to electron beam position. This leads mostly to preferential excitation of modes such as $D_1$ and $Q_1$, where charges are strongly localized on a triangular tip or the center of an edge (see Figure 4c), which intuitively explains the larger contribution of these modes in STEM experiments[25,28].

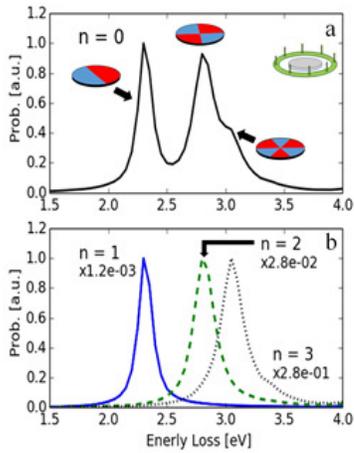

Figure 3. Calculated energy loss probability for generated on a silver disk of radius = 5 nm (aspect-ratio =10) at the centre of a vortex donut beam of radius 8 nm. a) Predicted spectra for the $n=0$ vortex (representing typical extended beam TEM measurement) where the peaks associated with the first multipoles can be easily recognized (dipolar, quadrupolar and hexapolar). b) When using vortex beams of order $n \geq 1$, just the SP mode matching the vortex order is excited, with several orders of magnitude intensity enhancement. The reinforcement factor goes from ~1000 to ~10 for vortex orders going form $n=1$ to 3. The displayed probabilities have been normalized in relation to the maximum of the spectra predicted for the donut-beam with order $n=0$.

As the surface modes of triangular NPs do not show such a clear in-plane rotational symmetry distribution (see Figure 4c), their match to the vortex beam spatial phase structure is weaker, then we should not expect such strong resonances. In addition, we must anticipate a certain impact parameter effect, as the distance particle edge to the electron donut beam is not uniform. In fact the triangle tips are always closer to the electron trajectories so dipolar modes ($D_1$, $D_2$) should present a more important contribution in relation to quadrupolar ones. But, in contrast to narrow electron beam experiments, modes such as $D_2$ and $Q_2$ gain importance when excited with a vortex beam because they display a somewhat better rotational symmetry (see Figure 4c).

Figure 4a,b compare energy loss probability expected for triangular particles located at the centre of a donut beam of different vortex orders ($n= 0$-$4$). The peak associated with dipolar modes (~2.3 eV) is the strongest for vortex orders 0, 1, 2, and 4; quadrupolar modes (~2.9 eV) gain importance for vortex orders 2 and 4. However for a vortex $n=3$ beam the hexapolar mode of the triangular NP can be resonantly excited at its intensity is 10-100 higher than the $D$ and $Q$ modes. We must note that in this situation the $H$ peak intensity is about three orders of magnitude smaller than the dipolar mode excited with vortex beam $n=0$ (representing somewhat typical experimental measurement).

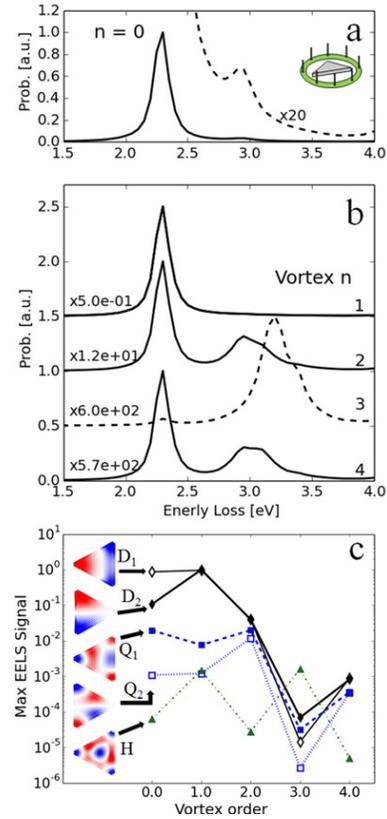

Figure 4. Energy loss probability estimated on silver triangle of side = 13.6 nm (aspect ratio =10) at the centre of a vortex donut beam of radius 10 nm. The displayed probabilities have been normalized in relation to the maximum of the spectra predicted for the donut-beam with order $n=0$. a) Predicted spectra for the $n=0$ vortex where we can observe the loss peaks associated with the surface mode pseudo-symmetry (dipolar at ~2.3 eV and quadrupolar at ~2.9 eV, see text for explanations). b) In contrast to spherical and disc-shaped particles the use vortex beams of order $n \geq 1$, the phase structure is not so effective to select/enhance the surface plasmon multipole, except for the dipolar and hexapolar modes. This is due to the complex interplay between particle symmetry and variable distances of the triangular nanoparticles edges to the annular electron beam; we must keep in mind that the dipolar modes are mostly excited for electron trajectories close to the tips, while quadrupolar modes are mostly excited

for electron trajectories near the centre of the triangles sides[25,26,28]. c) Comparison between the maximum intensity of the different surface modes (dipole $D_1$ and $D_2$, quadrupolar $Q_1$ and $Q_2$ and the hexapolar one $H$[28]) when excited using different vortex orders.

The precedent results derived for metal disks and triangles represent a clear manipulation of multipole selective excitation by choosing the proper vortex order beam. This is very promising, but we must keep in mind that these calculations have considered the perfect geometrical case of a particle at the vortex centre. Figure 5 shows the effect of placing a triangular NP out of the donut beam central point. As maybe expected, the phase structure resonant effect is reduced as a function of eccentricity. Although this results indicates that selection of higher multipole modes must follow stringent experimental requirements, it is interesting to note that dipolar effects (associated to $n=1$ vortex beams) are very robust. To a certain extent, this is very positive, because the different EELS spectral features (in the low and high loss energy range) are frequently compared with photon equivalent absorption spectroscopies using light[1], where dipolar selection rules apply. This is often the typical procedure in order to get a deeper understanding of materials electronic structure by comparing low and high spatial resolution results.

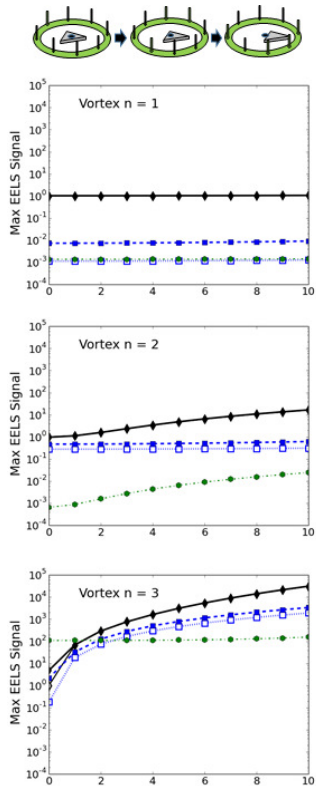

Figure 5. Comparison between the intensity of the different surface modes when placing the plasmonic nanoparticle in an eccentric position out of the centre of the donut beam, where the vortex spatial phase structure gets out of its perfect geometrical matching. The calculations have considered a triangular silver nanoparticle of side = 13.6 nm (aspect-ratio =10) at the centre of a vortex donut beam of radius 16 nm. The horizontal axis measures the eccentricity describing the progressive lateral shift of the triangle; the value=0 correspond to concentric situation, while eccentricity = 1 correspond to the shift where the particle edge is at the donut centre (see schematic draw at the top of the figure). Note that the weight of different modes changes significantly when the triangular particles is taken out of the donut beam, however no significant charges are observed for the vortex $n=1$.

All our previous theoretical studies have considered azimuthal phase effects on the excitation of surface plasmons on NPs located inside the donut beam. Another natural question arises: does this phase azimuthal structure generate measurable effects outside the donut beam? Probably due to partial coherence effects the phase information should extend outside the donut to a certain distance (a coherence length). To analyse this situation we have calculated energy loss line profiles for a focussed donut vortex beam (1 nm in diameter) scanning parallel to the side of a triangular metal particle (see Figure 6). For a beam without a phase structure ($n=0$) dipolar ($D$) modes are stronger for positions close to the triangles tips, while $Q$ modes are stronger close to centre of the sides, as already reported[25,26,28]. This tendency in reversed if the beam has $n=1$ vortex phase distribution; in this case the maximum of intensity is when the vortex beam is closer to inversion nodes of the surface charge density. In this configuration, the maximum interaction is attained because the different charge region contribution at each side of the node is reversed by the vortex phase. As an overall effect, this reduces the efficiency of the interaction (and the total spectrum intensity) by about three orders of magnitude.

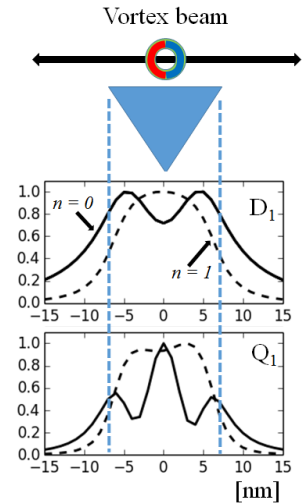

Figure 6. Comparison of intensity profile of surface modes $D_1$ and $Q_1$ expected when vortex beam scans a line parallel to a triangular nanoparticle side (triangle Side = 13.6 nm, aspect/ratio =10, donut-beam rad. = 1nm and, distance beam centre to triangle edge 2 nm). As expected for a normal focused electron beam, the $n=0$ vortex shows maxima of intensity close to triangle tips (indicates as dashed vertical lines) and side centre for the $D_1$ and $Q_1$ modes respectively. In con-

trast for a $n=1$ vortex, the maximum of intensity profile is located in front of the node of the surface charge density (see text for explanations). We have included the three-fold degenerate surface charge distribution for each of the studied modes.

Furthermore, vortex beams may be exploited to manipulate the coupling of dark surface modes in neighboring NPs[30]. When we place a standard electron beam between two nanoparticles or nanorods, we stimulate modes where the region closest to the electron trajectory should be polarized with positive charges generating a symmetrical mode[30]. If we place a vortex $n=1$ beam between the particles the phase structure will generate an antisymmetric configuration where the particles region closest to the vortex beam will have opposite charges on each particle. This mode is actually observed in optical absorption spectroscopy when the polarization is parallel to the nanorod axis[30] or on EELS studies if the beam is place far from the nanorod junction and close to the furthest nanorod tip. Also, manipulating the sign of charges in neighboring nanoparticles maybe used to control interparticle forces and can be exploited to manipulate NP coalescence.

In summary we have theoretically studied the interaction between the phase structure of electron vortex beam and nanoparticles surface plasmon modes for the case of non-penetrating trajectories. We have shown that vortex beam can be used to modify the weight of different SP mode excitation by matching the SP mode charge distribution azimuthal symmetry and the vortex beam order. The description of electron vortex beam developed here can be easily adapted for other existing BEM and discrete dipole approximation (DDA[31,32]) software packages. There is still a lot of work to be done to extend the present approximations to include retardation effects, penetrating trajectories and also consider the excitation of other energy loss mechanisms. This semi-classical approach just considers a perfect matching of symmetries, which in quantum mechanics means that the final state of the electron vortex wave should be the order $n=0$. Although, the theoretical results reported here do not represent a manipulation or control of selection rules, we expect that a more detailed full quantum mechanical description of vortex beam interaction will finally allow understanding of interaction and exploitation of phase structures in EELS studies in electron microscopy based spectroscopies.

As the phase structure of a vortex beam is azimuthal, the interaction with the NP becomes mostly dominated by interference terms based on the in-plane electric field components, while a conventional electron beam allows the study of modes which couple longitudinally to the electron trajectory. In this sense, electron vortex beams will allow the study of absorption anisotropy effects or access modes that can be considered as dark modes in EELS studies. This may also be extremely useful for studies associating tomography and low loss spectroscopy[33]. We must keep in mind that in a donut beam any rotational symmetry of the sample will always be in resonance with a matching azimuthal phase structure (any constant phase difference does not have any effect on the energy loss). When the electron beam has a spatial dephasing generated by multipolar lenses (ex. astigmatism[34]) it is necessary to get proper azimuthal angle alignment to induce resonances; however, this also opens the opportunity to select different preferential in-plane beam-sample interaction directions.

## ASSOCIATED CONTENT

**Supporting Information**. We present in detail the calculations associated with the introduction of donut-vortex beam phase structure into the BEM methodology. This material is available free of charge via the Internet at http://pubs.acs.org.

## AUTHOR INFORMATION

**Corresponding Author**

* E-mail: dmugarte@ifi.unicamp.br

## ACKNOWLEDGMENT

We are very grateful to M. Kociak for discussions on electron energy loss spectroscopy and the physics of surface plasmons. We acknowledge financial support from the Brazilian Agencies FAPESP (Grant. No. 2014/01045-0) and CNPq (Grant. No. 302767/2012-6 ). CD acknowledges funding from the ERC under grant number 259619 PHOTO EM.

# SUPPORTING INFORMATION:

# Controlling Multipolar Surface Plasmon Excitation through the Azimuthal Phase Structure of Electron Vortex Beams.

Daniel Ugarte[†,‡,*], Caterina Ducati[†].

[†]Department of Materials Science and Metallurgy, University of Cambridge, Cambridge CB3 0FS, United Kingdom.
[‡]Intituto de Física "Gleb Wataghin", Universidade Estadual de Campinas-UNICAMP, 13083-859 Campinas SP, Brazil.

**Content:**

In this Supporting Information we present in more detail the introduction of donut-vortex beam phase structure into the BEM method.

**Vortex Beam and BEM approach.**

Among the different approaches to calculate SP excitation by a high energy electron beam, the boundary element method (BEM), originally proposed by Garcia de Abajo *et al.* (1,2), is one of the most popular procedures. The BEM method can be easily implemented for homogeneous particles (described by a local dielectric response $\varepsilon_1(\omega)$, $\omega$ represent th energy) of arbitrary shape embedded in an uniform dielectric medium ($\varepsilon_2(\omega)$). The calculations require the knowledge of the surface charge distribution $\sigma(s)$ at an assumed abrupt interface between the two dielectric media. Using the quasi-static approximation (1,2), the surface charge $\sigma(s)$ is derived from Poisson's equations for a fast electron (descrived as $\Phi^{ext}$) traveling close to the dielectric particle. The BEM surface charge density is obtained from a self-consistent integral equation in energy domain (see details in (1)):

$$\Lambda(\omega)\, \sigma(s, \omega) \;=\; \mathbf{n}_s . \nabla \Phi^{ext}(s, \omega) + \int d s'\; F(s, s')\; \sigma(s', \omega)$$

where,

$$\Lambda(\omega) = 2\,\pi\, \frac{\varepsilon_2(\omega)+\varepsilon_1(\omega)}{\varepsilon_2(\omega)-\varepsilon_1(\omega)} \quad \text{and} \quad F(s, s') = -\frac{\mathbf{n}_s . (s - s')}{|s - s'|^3}$$

It is important to emphasize that the surface charge distribution can be built by the weighted contribution of different interface surface modes $\sigma^i(s, \omega)$, which are independent of energy and depend only on the shape to the considered nanoparticles (3). These so-called geometrical eigen-functions describe self-sustained surface charge oscillations modes (or surface plasmons, SPs), which are obtained from an eigenvalue ($\lambda_i$) integral equation:

$$2\,\pi\, \lambda_i\, \sigma^i(s) \;=\; \int d s'\; F(s, s')\; \sigma^i(s)$$

Ouyang and Isaacson (4) have shown the the $\lambda_i$ factors are real, and the modes form a complete basis set that satisfy a peculiar (bi-)orthogonality property:

$$\int d\mathbf{s} \int d\mathbf{s}' \frac{\sigma^i(\mathbf{s})\, \sigma^j(\mathbf{s}')^*}{|\mathbf{s} - \mathbf{s}'|} = \delta_{ij}$$

When the electron travels outside the particle volume (hereafter we will always assume this configuration), the final surface charge density is the weighted sum over the modes:

$$\sigma(\mathbf{s}, \omega) = \sum_i \frac{f_i(\omega)}{\varepsilon_2(\omega)\,[\,\Lambda(\omega) - 2\pi\lambda_i\,]} \sigma^i(\mathbf{s})$$

where $f_i(\omega)$ factors are (1):

$$f_i(\omega) = \int d\mathbf{s} \int d\mathbf{s}'\ \mathbf{n}_s \cdot \nabla \Phi^{ext}(\mathbf{s}, \omega) \frac{\sigma^i(\mathbf{s}')^*}{|\mathbf{s} - \mathbf{s}'|}$$

For an external electron beam, this expression can be further simplified using the second Green identity (5):

$$f_i(\omega) = 2\pi\,(2 + \lambda_i) \int d\mathbf{s}\ \sigma^i(\mathbf{s})^*\ \Phi^{ext}(\mathbf{s}, \omega)$$

Subsequently, the energy loss is calculated by considering the dissipative work performed on the fast electron by the electric field originated from the boundary polarization potential $\Phi^B$. Without loss of generality, we will consider an electron traveling along a rectilinear path, $\mathbf{r}_e = \mathbf{r}_0 + \mathbf{v}\,t$, where $\mathbf{v} = (0,0,-|v|)$ and $\mathbf{r}_0$ is perpendicular to the velocity $\mathbf{v}$. Finally, we consider that the medium around the particle is vaccum ($\varepsilon_2(\omega) \equiv 1$), then, the energy loss probability results:

$$P(\mathbf{r}_0, \omega) = \frac{1}{\pi v}\ Im\left\{-\Phi^B(\mathbf{r}_0, q_z, \omega)\big|_{q_z = \frac{\omega}{v}}\right\}$$

The boundary potencial can be expressed as a function of the surface charges (1,2):

$$\Phi^B(\mathbf{r}, \omega) = \sum_i \frac{f_i(\omega)}{\varepsilon_2(\omega)\,[\,\Lambda(\omega) - 2\pi\lambda_i\,]} \int d\mathbf{s} \frac{\sigma^i(\mathbf{s})}{|\mathbf{r} - \mathbf{s}|}$$

Then, the energy loss probability ($P^i$) due to each surface charge mode becomes (remind the the beam is external to the particle so there is no "Begrenzum effect" (1)):

$$P^i(\mathbf{r}_0, \omega) = \frac{1}{(1 + \lambda_i)\,\pi^2 v}\ Im\{-g_i(\omega)\ f_i(\omega)\ I_i(\mathbf{r}_0, \omega)\}$$

where $\qquad g_i(\omega) = \frac{2}{\varepsilon_1(\omega)\,(1+\lambda_i) + \varepsilon_2(\omega)\,(1-\lambda_i)}$

and $\qquad I_i(\mathbf{r}_0, \omega) = \int d\mathbf{s}\ \sigma(\mathbf{s})\, e^{-i\frac{s^\| \omega}{v}}\ K_0\left(\frac{\omega\,|\mathbf{r}_0 - \mathbf{s}^\perp|}{v}\right)$

The indexes $\perp$ ($\|$) represents the vector component perdendicular (parallel) to the electron trajectory, and $K_0$ is the modified Besssel function of order 0.

It is important to note that for non penetrating trajectories $f_i(\omega) = 2\pi\,(2+\lambda_i)\,I_i^*$, what leads to a very simple expression (3,6), where the main energy dependence is contained in the $g_i(\omega)$ factor, and geometrical aspects in a $\|I_i\|^2$ factor:

$$P_n^i(\boldsymbol{r_0}, \omega) = \frac{4}{\pi v^2} \frac{1}{(1 + \lambda_i)} \; Im\{-g_i(\omega)\} \; \|I_i(\boldsymbol{r_{0,k}}, \omega)\|^2$$

In order to study the excitation of SP by electron vortex beams (7-9), we must find a physical description that can be used to extend the BEM formalism. Mohammadi *et al.* (10) have considered an electron travelling in a helical trajectory to generate an electron beam with angular momentum in order to study magnetic responses. In contrast, we aim is to find a description where the azimuthal phase structure (11-13) of the vortex beam could be taken into account within a quasi-static electromagnetic calculation.

Basically a vortex beam of order *n* can be described in cylindrical coordinates by the following wave function (11):

$$\psi_{vortex} = f(r) e^{in\varphi} \, e^{ik_z z}$$

Vortex beams present a central region with vanishing intensity, then we may approximate the beam cross section described by *f(r)* as being an infinitively thin ring of radius $R_o$. In analogy to the numerical implementation in the BEM approach, we discretize the vortex beam as an ensemble of non-interacting linear electron trajectories; and, in order to generate the azimuthal phase, we add an additional spatial phase term *exp(inφ)* for each position along the azimuthal $\varphi$ angle ($\boldsymbol{r_{0,k}} = \boldsymbol{r_0}(\varphi_k), |\boldsymbol{r_{0,k}}| = R_0$ ). These approximations lead to the following the electron density:

$$\rho(\boldsymbol{r}) = \sum_k e^{i n \varphi_k} \, \delta(\boldsymbol{r} - \boldsymbol{r_{0,k}})$$

Subsequently, we will calculate the energy loss adding the contribution of each electron beam considering the vortex phase factor. The calculation of the energy loss considering this additional summatory does not add major additional difficulties to the dielectric formalism:

$$\Delta E = \int_{-\infty}^{\infty} \rho^* \; E_z^B \; dz$$

$$\Delta E = \sum_k e^{-i n \varphi_k} \int_{-\infty}^{\infty} E_z^B(\boldsymbol{r}(\varphi_k)) \, dz$$

Within this pseudo-static calculation the dissipation work results from the electric field ($E_z^B$) due to the boundary potential evaluated at the electron position. As the BEM surface charge density generating the boundary potential $\Phi^B$ is derived considering the whole ring-shaped beam, the calculation will result in a weighed sum including different spatial phase factors $e^{-i n \varphi_k}$. This leads to the typical result of the BEM approach with an additional summatory considering the beam along the ring-shaped beam circumference. The probability of energy loss for the mode *i* when using a vortex of order *n* becomes:

$$P_n^i(\boldsymbol{r_0}, \omega) = \frac{1}{(1 + \lambda_i) \, \pi^2 v} \; Im\left\{-g_i(\omega) \; f_i(\omega) \sum_k e^{-i n \varphi_k} \, I_i(\boldsymbol{r_{0,k}}, \omega)\right\}$$

where $f_i(\omega)$ is:

$$f_i(\omega) = \frac{4\pi}{v} \; (2 + \lambda_i) \sum_k e^{i n \varphi_k} \, I_i^*(\boldsymbol{r_{0,k}}, \omega)$$

This yields:

$$P_n^i(\boldsymbol{r_0}, \omega) = \frac{4}{\pi v^2} \frac{(2 + \lambda_i)}{(1 + \lambda_i)} \; Im\{-g_i(\omega)\} \; \left\| \sum_k e^{-i n \varphi_k} I_i(\boldsymbol{r_{0,k}}, \omega) \right\|^2$$

This expression is almost identical to the one derived for an individual electron trajectory, but each azimuthal position of the discretized electron beam contributes independently to the energy loss through the $I_i(\boldsymbol{r_{0,k}}, \omega)$ integral. Subsequently, these terms are added taking into account the spatial phase factor and, then the square modulus is evaluated. This classical calculation (introducing the phase factor) seems intuitively correct; in fact, the expected result for a semi-classical quantum calculation considering phase related interference effects between each electron trajectory.

In this way, the vortex phase structure will directly interact with azimuthal symmetry of SP modes described by $\sigma^i(\boldsymbol{s})$, what can be described in a full quantum calculation as taking the final state a having a vortex order $n=0$ (see Supplementary Material associated with Ref. 14). We have implemented the non-retarded BEM methodology considering the phase effects in Python language, and the singular integrals were evaluated following the suggestion reported by Hohenester and Trugler (15). In the work presented here, we will apply this approach to analyse the energy losses associated with SP modes for particles of different morphologies located at the centre of the extended vortex beam and, also, considered focused vortex beams where the plasmonic nanoparticles are located outside the donut-beam.